\documentstyle[11pt,newpasp,twoside]{article}
\def\edcomment#1{\iffalse\marginpar{\raggedright\sl#1\/}\else\relax\fi}
\reversemarginpar

\begin{document}
\title{Transport Phenomena and Light Element Abundances in the Sun and Solar Type Stars}
 \author{Sylvie Vauclair}
\affil{Laboratoire d'Astrophysique, 14 av. Ed. Belin, 31400-Toulouse, France }

\begin{abstract}

 The observations of light elements in the Sun and Solar type stars give special clues for
understanding the hydrodynamical processes at work in stellar interiors. In the Sun $^7$Li
is depleted by 140 while $^3$He has not increased by more than  $\cong~10\%$ in 3 Gyrs.
Meanwhile the inversion of helioseismic modes lead to a precision on the sound velocity of
about $.1\%$. The mixing processes below the solar convection zone are constrained
by these observations. 
Lithium is depleted in most Pop~I solar type stars. 
In halo stars however, the lithium abundance seems constant in the ``spite
plateau" with no observed dispersion, which is difficult to reconcile with the theory of
diffusion processes. 
In the present paper, the various relevant observations will be
discussed. It will be
shown that the $\displaystyle \mu $-gradients induced 
by element settling may help solving the ``lithium
paradox".

\end{abstract}

\section{Introduction}
Element diffusion and mixing processes in stellar interiors are now widely constrained,
first by detailed observations of abundances, second by helio and asteroseismology.
In most cases however, pure microscopic diffusion in stars would lead to abundance
variations much larger than those observed : mild macroscopic motions in stellar radiative
zones are definitely needed to account for the observations. This gives strong constraints
on the kind of mixing processes allowed. Other constraints come from the consequences of the
nuclear reactions occuring in stellar interiors : in some cases stellar mixing from the
atmosphere down to the regions of nuclear processing is needed to explain the observed
element abundances. This is the case, for example, to account for the depletion of lithium
in the Sun and solar type stars. 

Lithium observations in main-sequence population I field stars and galactic clusters show a
large abundance dispersion which has been extensively studied in the literature (see
 reviews by Deliyannis 2000, Charbonnel 2000, Michaud 2000 and Pinsonneault 2000). 
The lithium abundance
decreases for decreasing effective temperature below 5500K and the depletion increases with
increasing age. This is generally attributed to the deepening of the convective zone,
associated with some mild mixing process connecting the bottom of the convective zone
with the nuclear destruction region.

Lithium is also depleted in F-type
stars (the so-called ``Boesgaard dip"). Several possible reasons have been invoked to explain
this feature, most related to mixing and nuclear destruction.
Element segregation has been proved negligible here as it would lead to
unobserved variations of metal abundances (Turcotte et al 1998)
and beryllium (Boesgaard 2000).

On the other hand, observations of lithium in main-sequence population~II field stars show
remarkably constant abundances, with a very small dispersion (e.g. Bonifacio and Molaro
1997)
Why is lithium destroyed in Pop~I stars while it does not seem destroyed in Pop~II stars?

For the same effective temperatures, the convective zone is smaller in Pop~II stars than in
Pop~I stars because of their smaller metallicity. Meanwhile they have a smaller rotation
velocity on the average. This could explain why the lithium destruction induced by nuclear
reactions is smaller in these stars than in Pop~I stars.
However the element segregation is more important for smaller
densities and smaller rotation, so that this process should lead to a visible lithium
depletion, which is not observed (Vauclair and Charbonnel 1995 and 1998).
This represents the so-called
``lithium paradox". Here we suggest that the influence of
$\displaystyle \mu $-gradients on the rotation-induced mixing may help solving this paradox.

\section{Competition between rotation induced mixing and element diffusion}

In rotating stars, the equipotentials of  ``effective gravity'' (including the centrifugal
acceleration) have
ellipsoidal shapes while the energy transport still occurs in a spherically
symetrical way. The resulting thermal imbalance must be compensated
by macroscopic motions: the so-called ``meridional circulation''
(Von Zeipel 1924). The stellar regions outside the convective
zones cannot be in complete radiative equilibrium. They
are subject to entropy variations given by~:
\begin{eqnarray}
\rho T \left( {\partial S \over \partial t} + {\bf u} \cdot {\bf \nabla} S\right)
 & = & - {\bf \nabla} \cdot {\bf F} +
\rho  \varepsilon _{n} \nonumber \\
& = & \rho  \varepsilon _{\Omega} \ (\not= 0)
\end{eqnarray}
where $\displaystyle {\bf F}$ represents the heat flux, $\displaystyle \varepsilon _{n}$ the
nuclear
energy production and $\displaystyle \varepsilon _{\Omega}$ an energy generation
rate which results from sources and sinks of energy along the
 equipotentials.

The vertical component of the meridional velocity $\displaystyle u_{r}$ is computed
as a function of $\displaystyle \varepsilon _{\Omega}$ in the stationary
regime (from eq. 1):

\begin{equation}
u_{r} = \left( {P \over C_{p}\rho T }\right)
{\varepsilon _{\Omega} \over g }
\end{equation}
which, for a perfect gas, reduces to:
\begin{equation}
u_{r} =
{\varepsilon _{\Omega} \over g} \
{{\bf \nabla} _{{\rm ad}}
 \over{\bf \nabla}_{{\rm ad}} - {\bf \nabla} + \nabla_{\mu }}
\end{equation}
where $\displaystyle g$ represents the local gravity,
$\displaystyle {\bf \nabla}_{{\rm ad}}$ and
$\displaystyle {\bf \nabla}$ the usual adiabatic and real ratios
$\displaystyle \left( {d \ln T\over d \ln P }\right)$
and $\displaystyle \nabla_{\mu }$ the mean molecular weight
contribution
$\displaystyle \left( {d \ln \mu  \over d \ln P}
\right)$.

The expression of $\displaystyle \varepsilon _{\Omega}$ is
computed by expanding the right-hand-side of eq.~(1) on a level surface
and writing that its mean value vanishes.

Mestel (1953, 1957 and 1965)
pointed out that, in the presence of vertical $\displaystyle \mu $-gradients,
$\displaystyle \varepsilon _{\Omega}$ contains two kinds of terms : those related
to the resulting horizontal variations of $\displaystyle \mu $:
the so-called ``$\displaystyle \mu $-induced currents''
$\displaystyle  E_{\mu }$
and those independent of
$\displaystyle \mu $, the so-called ``$\displaystyle \Omega $-induced currents''
$\displaystyle E_{\Omega}$ .
The expression of $\displaystyle \varepsilon _{\Omega}$ obtained
in this case has been
derived in detail by Maeder and Zahn (1998), who took into account several effects which
were not
included in the previous computations: more general equations
of state instead of perfect gas law, presence of a thermal
flux induced by horizontal turbulence,
non-stationary cases.

Vauclair (1999) discussed more simple expressions, valid only for negligible differential
rotation. In this case
$\displaystyle \mu $-currents are opposite
to $\displaystyle \Omega$-currents in most of the star 
and $\displaystyle \varepsilon _{\Omega}$  may be written :

\begin{equation}
\varepsilon _{\Omega} =
\left( {L \over M}\right)
\left( E_{\Omega} + E_{\mu }\right) P_{2}
(\cos \theta)
\end{equation}
with:
\begin{eqnarray}{}
E_{\Omega} & = & {8 \over 3}
\left({\Omega^{2}r^{3} \over GM }\right)
\left( 1 - {\Omega^{2} \over 2\pi G\overline \rho  }
\right) \\
E_{\mu } & = &  {\rho _{m} \over \overline \rho  }
\left\{
{ r \over 3 } \
{d \over dr }
\left[
\left(
H_{T}
{d \Lambda \over dr}\right)
- (\chi_{\mu } + \chi_{T} + 1) \Lambda \right]
- {2 H_{T} \Lambda \over r } \right\}
\end{eqnarray}

Here $\displaystyle \overline \rho $ represents
the density average on the level surface
$\displaystyle (\simeq \rho )$ while
$\displaystyle \rho _{m}$ is the mean density inside
the sphere of radius $\displaystyle r$;
$\displaystyle H_{T}$ is the
temperature scale height;
$\displaystyle \Lambda$  represents the
horizontal $\displaystyle \mu $ fluctuations
$\displaystyle {\tilde{ \mu}\over \overline \mu } $;
$\displaystyle \chi _{\mu }$ and
$\displaystyle \chi _{T}$ represent the
derivatives:
\begin{equation}
\chi_{\mu } =
\left(
{\partial \ln \chi \over \partial \ln \mu  }\right)_{P,T}
\quad  ; \quad
\chi_{T} =
\left( {\partial \ln \chi \over \partial \ln T }\right)_{P, \mu }
\end{equation}

Vertical $\displaystyle \mu $-gradients may occur in
stars due to two different processes : first the nuclear reactions which occur in the
stellar cores, second the helium settling which occurs in the outer layers.
The importance of the first process
in reducing or even suppressing the meridional motions has been demonstrated
several times in the literature (e.g. Huppert and
Spiegel 1977). The second process on the other hand has not been extensively studied.
We claim here that it may play a crucial role for understanding the lithium problem in Pop~I
and Pop
~II stars.

\section{Application to Pop~II stars}

Computations of $\displaystyle \mu $-currents induced by the helium settling in halo stars
have been performed by Vauclair 1999 and Th\'eado and Vauclair 2000 a and b.
We found that, for
slow rotation, $\displaystyle \mu $-currents cancel
$\displaystyle \Omega$-currents for very small concentration gradients,
corresponding to  $\displaystyle \mu $-gradients
of order $\displaystyle 10^{-15}$~cm$^{-1}$.

Let us summarize the situation of a slowly rotating star
in which element settling leads to an increase of the
$\displaystyle \mu $-gradient below the outer convection
zone.
At the beginning, the star is homogeneous and meridional
circulation can occur, leading to upward flows in the polar
regions and downward flows in the equatorial parts
(except in the very outer layers where the Gratton-\"Opik
term becomes important, which we do not discuss here).
The $\displaystyle \mu $-currents, opposite to the
classical $\displaystyle \Omega$-currents, are first
negligible. The $\displaystyle \mu $-gradients
increasing with time because of helium settling, the order
of magnitude of the $\displaystyle \mu $-currents also
increases until it reaches 
the value for which the circulation vanishes.

This does not occur all at once:
as the  $\displaystyle \mu $-gradient decreases
with depth below the convective zone, we expect that the meridional
circulation freezes out step by step (see figure 1 of Th\'eado
and Vauclair 2000a). An equilibrium situation
may be reached, in which the temperature and mean molecular
weight gradients along the level surfaces are such that
$\displaystyle \Omega$-currents and
$\displaystyle \mu $-currents cancel each other.

Once it is reached, this equilibrium situation is quite robust. Suppose that some
mechanism leads to a decrease of the
 $\displaystyle \mu $-gradient: then $\displaystyle \vert E_{\mu }\vert$ becomes smaller
than $\displaystyle \vert E_{\Omega }\vert $ and the circulation
tends to be restablished in the $\displaystyle \vert E_{\Omega }\vert $
direction, thereby restoring the original $\displaystyle \mu $ gradient.
Suppose now that the
 $\displaystyle \mu $-gradient is increased.
Then $\displaystyle \vert E_{\mu }\vert$ becomes larger than
$\displaystyle \vert E_{\Omega }\vert $ and the circulation begins in
the $\displaystyle E_{\mu }$ direction.
Here again the original gradient is restored.

When the meridional circulation is frozen below the convective zone, helium settling could 
proceed further; however, due to the increase of the diffusion time scale with depth,
this would modify the  $\displaystyle \mu $-gradient. We may thus expect that
$\displaystyle \mu $-currents would take place and restore the original equilibrium
gradient, thereby strongly reducing the microscopic diffusion
(Th\'eado and Vauclair 2000b).
This self-regulating process could be the reason for the low dispersion of the
lithium abundance in the lithium plateau of halo stars.

\section{Discussion : Pop I versus Pop II stars}

There are many observations in stars which give evidences of mixing processes
occuring below the outer convective zones as, for example, the lithium depletion
observed in the Sun and in galactic clusters. The process we have described
above should not apply in all these stars. The reason could be related to the
rapid rotation of young stars on the ZAMS and to their subsequent
 rotational braking.

The abundance determinations in the solar photosphere show that lithium
has been depleted by a factor of about 140 compared to the protosolar
value while beryllium has not been
depleted by more than a factor 2, and maybe much less, as discussed by Balachandran and
Bell (1997).
These values represent strong constraints on the mixing processes in the solar interior.

Observations of the $^3$He/$^4$He ratio in the solar
wind
and in the lunar rocks (Geiss 1993,
Geiss and Gloecker 1998) show that this ratio may not
have increased by more than $\cong~10\%$ since 3 Gyr in the Sun.
While the occurence of some mild mixing below
the solar convective zone is needed to explain
the lithium depletion ,
the $^3$He/$^4$He
observations put a strict constraint on its efficiency. The only way to obtain such a result
is to postulate a mild mixing, which would be efficient down to the lithium nuclear burning
region but not too far below, to preserve the original $^3$He abundance. The efficiency of
this mixing should also decrease with time, as the $^3$He peak itself builts up during the
solar life.

It is interesting to compute the minimum
enhancement of the $^3$He/$^4$He ratio implied by the lithium observed
depletion.
Vauclair and Richard 1998 showed that it is
possible to deplete lithium by a factor larger than $100$ as observed
and
not increase $^3$He/$^4$He by more than 5 percent since the solar
origin.  In this case beryllium is only depleted by about 10 percent.

Such a confined mixing zone is also needed from helioseismology~:
although the introduction of pure element settling in the solar models considerably
improves the consistency with the seismic Sun, some discrepancies do remain, particularly
below the
convective zone where a "spike" appears in the sound velocity (Richard et al 1996,
Turck-Chi\`eze et al. 1998). It has been shown that this behavior may be due to the helium
gradient which would be too strong in case of pure settling. Mild macroscopic motions below
the convective zone slightly decrease this gradient and helps reducing the discrepancy
(Richard et al 1996, Corbard et al 1998, Brun et al 1998).
The helium profiles directly obtained from
helioseismology (Basu 1998, Antia and Chitre 1998) show indeed a helium gradient smoother
than the gradient obtained with pure settling.

The constraints implied by both the helioseismic inversions and
abundance determinations in the Sun converge towards the existence of a
small mild mixing region below the convective zone, which would extend
down to a depth of the order of one scale height.
The implied mixing region must be very mild, with diffusion
coefficients of $10^3$ - $10^4$ only.
It must also be completely deconnected
from the solar core. No mixing can indeed be allowed down to the nuclear
energy production region as it would lead to a sound velocity
incompatible with helioseismology. In particular the mixing processes
invoked by Morel and Schatzman 1996 to decrease the neutrino fluxes are excluded by
helioseismology (Richard and Vauclair 1997).

Mixing processes localized at the boundary between convective and radiative regions include
overshooting and regions of large differential rotation like the ``tachocline" below the
solar convective zone. Up to now, overshooting was generally treated in the models simply as
a continuation of the convective zone on a fraction of a pressure scale height. Recent
parametrisations use a diffusion coefficient which decreases exponentially with decreasing
radius (Freytag et al 1996). The tachocline, which represents in the present Sun the small
boundary between the region of large differential rotation (in the convective zone) and the
region of solid rotation (in the radiative zone below) is also treated as a mixed layer with
an exponentially decreasing diffusion coefficient (Brun et al 1998, Richard 1999). Results
are encouraging, although more sophisticated numerical simulation including 2-D abundance
variations would be needed to go further.

In any case, the self-regulating process that we have discussed for halo stars in section~3
would not apply below the convective zone in the Sun and solar type stars because of the
differential rotation which takes place there. Such a differential rotation would not be
expected in halo stars if we suppose that they always rotated slowly and thus did not suffer
large transport of angular momentum. The different behavior for the lithium abundance in
Pop~I and Pop~II stars could thus be directly related to their rotation history.

\end{document}